\documentstyle[pra,aps,preprint]{revtex}
\tightenlines

\newcommand{\subsub}[3]{{#1_{\hskip-1pt\scriptscriptstyle #2}}_{{ }_{\hskip-2pt \scriptscriptstyle #3}}\hskip0pt}

\newcommand{\ii}{{\rm i}}

\newcommand{\fett}[1]{\mbox{\boldmath $#1$}}

\newcommand{\dd}{\hskip -4pt {\rm d}\hskip-1pt^{{ }^3}\hskip -4pt}

\begin{document} 
\draft 
\title
{$\fett N$-Nucleon Effective Generators of the Poincar\'e Group
Derived from a Field Theory}
\author
{
A. Kr\"uger and W. Gl\"ockle\\
{\it Institut f\"ur Theoretische Physik II, Ruhr-Universit\"at Bochum,\\
D-44780 Bochum, Germany}
}
\maketitle

\begin{abstract}
It is shown that the ten generators of the Poincar\'e group acting in
the Fock space of nucleons and mesons and based on standard
Lagrangians can be blockdiagonalized by one and the same unitary
transformation such that the space of a fixed number of nucleons is
separated from the rest of the Fock space. The existence proof is
carried through in a formal power series expansion in the coupling
constant to all orders. In this manner one
arrives at effective generators of the
Poincar\'e group which act in the two subspaces separately. 
\end{abstract}

\section{Introduction}
Low energy nuclear physics below the pion threshold is naturally
formulated in terms of a fixed number of nucleonic degrees of
freedom. In the overwhelming number of applications a nonrelativistic
framework is used. This however is not sufficient if one investigates
for instance electron scattering with high three-momentum transfers as
one encounters in typical experiments performed nowadays. 
Also it is still an open question whether relativistic effects
play a significant role when calculating the binding energy of
nuclei. In three-nucleon scattering it has been found recently
\cite{abfalterer}
that the total $nd$ cross section evaluated with most modern $NN$
forces and based on rigorous solutions of the $3N$ Faddeev equations
deviate from the data above $\approx 100$ MeV nucleon laboratory
energy. That discrepancy reaches about $10\%$ at $300$ MeV and is
very likely caused by the neglect of relativity. On all these grounds
a relativistic generalization of the usual Schr\"odinger equation for
$N$ interacting nucleons is highly desirable.

In \cite{dirac} Dirac proposed three forms of relativistic quantum
mechanics for a given number of interacting particles. A realization
thereof in the instant form was given by Bakamjian and Thomas
\cite{bakthom}. That scheme however violates cluster separability
\cite{foldy}. Being less ambitious and searching just for
relativistic correction terms
to the generators of the Galilean group in leading orders
Foldy and Krajcik have discussed \cite{folkra}  a
$1\over{\rm c}^2$ expansion of the ten generators of the Poincar\'e
group. This scheme has been
applied recently in a realistic context in the $3N$ system \cite{carlson}. A
way to treat the defect in the Bakamjian and Thomas scheme with respect to the
cluster separability has been found by Sokolov \cite{soko} and also worked
out by Coester and Polyzou \cite{copo}. An extensive overview over the
whole subject is given in \cite{keipo}.

There is however also another approach to the generators of the
Poincar\'e group for a fixed number of particles. Relativistic field
theory provides generators which act in the full Fock space. Thinking
of applications for nuclear physics one considers interacting fields
of nucleons and mesons. To arrive at generators which act in the space
of a fixed number of $N$ nucleons one has to eliminate the mesonic
degrees of freedom as well as the ones for antiparticles. A way to do
this has been proposed in \cite{glmu} and worked out in lowest order in
the coupling constant for a field theory of ``scalar nucleons''
interacting with a scalar meson field. While this has been formulated
in the instant form a corresponding derivation can also be performed
in the light front form \cite{muller1}. Numerical investigations based on
those effective generators determined in leading order in the coupling
constant have been carried through in \cite{muller2}, \cite{glno} and
\cite{hagl}.

In this article we want to show that the derivation proposed in
\cite{glmu} can be carried through to arbitrary order in the coupling
constant. Thus the effective generators of the Poincar\'e group in an
$N$ nucleon subspace do exist at least in the sense of a formal series
expansion. It will be interesting to investigate whether those
generators are automatically also cluster separable. This is left to a
future study.

In section II we formulate our way to derive the effective generators in
the $N$ nucleon subspace out of a field theoretical model of
interacting nucleon and meson fields. The existence proof is carried
through in section III. We summarize briefly in section IV.

\section{Conditions for the effective generators}
We consider a field theory of interacting nucleons, antinucleons, and
mesons given by a Lagrangian of the form
 
\begin{equation}
{\cal L}={\cal L}_0+{\cal L}_I\label{lagrange}
\end{equation}
where ${\cal L}_0$ is the free part and the interacting part ${\cal
L}_I$ is linear in the coupling constant $g$. We also  assume
that ${\cal L}_I$ is linear in creation and absorption operators for
mesons as is
the case for often used field theories. In a standard manner \cite{weinberg}
one arrives at the ten generators of the Poincar\'e group for constant
time slices (instant form). The Hamiltonian and the three boost
operators carry interactions, whereas the
total  momentum and angular
momentum operators are the free ones. The latter two leave the plane
$t=$const invariant. Thus one has in obvious notation
\begin{eqnarray}
H	&=&	H_0+H_I\label{HH}\\
K_i	&=&	{K_0}_i+{K_I}_i\label{KK}\\
P_i	&=&	{P_0}_i\label{PP}\\
J_i	&=&	{J_0}_i\label{JJ}
\end{eqnarray}
where due to Eq. (\ref{lagrange})\\
\parbox{12cm}
{
\begin{eqnarray*}
H_I&\sim&g\\
{K_I}_i&\sim&g
\end{eqnarray*}
}
\hfill
\parbox{12mm}
{
\begin{equation}
\label{HHa}
\end{equation}
}\\
These ten operators fulfill the Lie algebra of the Poincar\'e group:
\begin{eqnarray}
&&[{ P}_{i},H]=0\label{lie1}\\
&&[{ J}_{i},H]=0\label{lie2}\\
&&[{ P}_{i},{ P}_{j}]=0\label{lie3}\\
&&[{ J}_{i},{ J}_{j}]=\ii\epsilon_{ijk}{ J}_{k}\label{lie4}\\
&&[{ J}_{i},{ P}_{j}]=\ii\epsilon_{ijk}{ P}_{k}\label{lie5}\\
&&[{ J}_{i},{ K}_{j}]=\ii\epsilon_{ijk}{ K}_{k}\label{lie6}\\
&&[H,{ K}_{i}]=-\ii { P}_{i}\label{lie7}\\
&&[{ K}_{i},{ K}_{j}]=-\ii\epsilon_{ijk}{ J}_{k}\label{lie8}\\
&&[{ P}_{i},{ K}_{j}]=-\ii\delta_{ij}H\label{lie9}
\end{eqnarray}
Formally one can verify that using the equal time commutation
relations of the underlying fields. Because ${\cal L}_I$
is assumed to be linear in the creation and annihilation operators
for mesons, $H_I$ and ${K_I}_i$ will be linear in these operators too.
Hence the eigenstates of $H$ will necessarily contain an
infinite number of mesons in addition to the nucleons (and
antinucleons). The behaviour of such an eigenstate under Lorentz
transformation, however, is transparent. We regard the operator of
four momentum 
\begin{equation}
{P}^\mu\equiv(H,P_1,P_2,P_3)\label{Pmu}
\end{equation}
and consider a Lorentz transformation $T(\Lambda,a)$ defined by
\begin{equation}
x^\mu{\buildrel{T}\over \longrightarrow}
x'^\mu={\Lambda^\mu}_\nu x^\nu+a^\mu\label{LT}
\end{equation}
Related to $T$ is a unitary operator $U(\Lambda,a)$ acting in the
Hilbert space spanned by the eigenstates of $H$. A consequence of the
commutation relations (\ref{lie1})-(\ref{lie9}) are the transformation
properties of $P^\mu$:
\begin{equation}
{P}^\mu{\buildrel{T}\over\longrightarrow}
{P'}^\mu=U{P}^\mu U^\dagger={\Lambda_\nu}^\mu {P}^\nu
\label{trans}
\end{equation}
Because of Eqs. (\ref{lie1}) and (\ref{lie3}) there exist simultaneous
eigenstates related to the four components of the four-momentum operator,
which fulfill
\begin{equation}
{P}^\mu|\Psi_p\rangle=p^\mu|\Psi_p\rangle\label{allgschr}
\end{equation}
Applying $U$ and using Eq. (\ref{trans}) one gets
\begin{equation}
{\Lambda_\nu}^\mu{P}^\nu U|\Psi_p\rangle	=	p^\mu
U|\Psi_p\rangle\label{step2}
\end{equation}
This can be rewritten into
\begin{equation}
{P}^\nu U|\Psi_p\rangle	=	{\Lambda^\nu}_\mu p^\mu 
					U|\Psi_p\rangle	
\label{step3}
\end{equation}
Thus up to a phase factor we get
\begin{equation}
U|\Psi_p\rangle	=|\Psi_{\Lambda p}\rangle
\label{geboostet}
\end{equation}
and the ``four dimensional Schr\"odinger equation'' (\ref{allgschr})
reads in the new frame of reference
\begin{equation}
{P}^\mu |\Psi_{\Lambda p}\rangle=(\Lambda p)^\mu|\Psi_{\Lambda
p}\rangle
\label{allgschr1}
\end{equation}
Therefore the simultaneous eigenstates of $P^\mu$ in the new frame are
eigenstates in the old frame with Lorentz transformed eigenvalues of
the overall four momentum.

We pose now the question if one can find a representation of the
Poincar\'e algebra being restricted to a subspace of the Fock space
with a fixed number of particles. We want to call those generators of
the Poincar\'e group ``effective''. If it is possible to find an
effective representation of the Poincar\'e group one is able to
formulate an effective Schr\"odinger equation in the subspace of a
given number of particles, say $N$ nucleons and no mesons, very much alike
(\ref{allgschr}). The interesting point about that is that this
equation would be easier to solve than Eq. (\ref{allgschr}) since the number
of degrees of freedom is finite now. In addition, since we assume the
Poincar\'e algebra to be fulfilled in that subspace, this effective
Schr\"odinger equation inherits the nice transformation properties of Eq.
(\ref{allgschr1}). 

A way to find effective generators is to unitarily
transform the generators (\ref{HH})-(\ref{JJ}) by an operator $\cal U$
such that all ten generators are put into a blockdiagonal shape at the
same time. One block would refer to the $N$ nucleon subspace, the
other block to the rest and the two blocks would not be coupled. Under
a unitary transformation the commutation relations remain valid, of
course.
Let us denote the projection on the subspace of $N$ nucleons by $\eta$
and the projection on the rest by $\Lambda\equiv 1-\eta$. Then what we
are looking for is a unitary transformation of the form 
\begin{eqnarray}
H&{\buildrel{\cal U}\over\longrightarrow}& \tilde H=\eta\tilde
H\eta+\Lambda\tilde H\Lambda\label{Htilde}\\
{K_i}&{\buildrel{\cal U}\over\longrightarrow}& \tilde {K_i}=\eta\tilde
{K_i}\eta+\Lambda\tilde {K_i}\Lambda\label{Ktilde}\\
{P_i}&{\buildrel{\cal U}\over\longrightarrow}& \tilde {P_i}=\eta\tilde
{P_i}\eta+\Lambda\tilde {P_i}\Lambda\label{Ptilde}\\
{J_i}&{\buildrel{\cal U}\over\longrightarrow}& \tilde {J_i}=\eta\tilde
{J_i}\eta+\Lambda\tilde {J_i}\Lambda\label{Jtilde}
\end{eqnarray}
While $H$ and $K_i$ (in the instant form) couple the $\eta$ and
$\Lambda$ spaces, this is by assumption no longer the case for
$\tilde H$ and $\tilde K_i$ and the operators $\eta\tilde H\eta$,
$\eta\tilde K_i\eta$, $\eta\tilde P_i\eta$ and $\eta\tilde J_i\eta$
are effective generators of the Poincar\'e group.
Now one can look for eigenstates of
$\tilde P^\mu$ whose $\Lambda$-components are zero. Lorentz
transformations on those states are generated by the effective
operators and we may write down the effective Schr\"odinger equation
\begin{equation}
\eta\tilde P^\mu\eta|\psi\rangle=p^\mu\eta|\psi\rangle
\label{allgschr2}
\end{equation}
In \cite{glmu} such a path has been initiated and will be worked out more
stringently now. In \cite{okubo} Okubo proposed a way to transform an
arbitrary hermitian operator
\begin{equation}
O=\left(\matrix{\eta O\eta&\eta O\Lambda\cr
		\Lambda O\eta&\Lambda O\Lambda}\right)\label{okuboso}
\end{equation}
into a block diagonal form by means of a unitary transformation $\cal U$:
\begin{equation} 
O\longrightarrow\tilde O={\cal U}O{\cal U}^\dagger=\eta \tilde O\eta+\Lambda \tilde O\Lambda
\label{okubo1}
\end{equation}
We follow Okubo for the choice of the unitary operator
\begin{equation}
{\cal U}=\pmatrix{\eta {\cal U}\eta & \eta {\cal U}\Lambda\cr
	\Lambda {\cal U}\eta & \Lambda {\cal U}\Lambda}
=\pmatrix{(1+A^\dagger A)^{^{-{1\over2}}}\eta & (1+A^\dagger
	A)^{^{-{1\over2}}}A^\dagger \cr
	-(1+A A^\dagger)^{^{-{1\over2}}}A & (1+A
	A^\dagger)^{^{-{1\over2}}}\Lambda}
\label{oktrans}
\end{equation}
where $A$ has the form
\begin{equation}
A=\Lambda A\eta\label{A}
\end{equation}
Unitary transformations within the subspaces $\eta$ and $\Lambda$ are
put to $1$. Using the forms (\ref{okuboso}) and (\ref{oktrans})
the requirement for blockdiagonalization is
\begin{equation}
\Lambda \biggl({[}O,A]+O-AOA\biggr)\eta =	0
\end{equation}
Since it is \`a priori not obvious that it will be possible to
blockdiagonalize each generator using the same $\cal U$ we label $A$ with
the generator to be blockdiagonalized. Noting Eqs. (\ref{PP}) and
(\ref{JJ})
telling that $P_i$ and $J_i$ do not connect the $\eta$ and
$\Lambda$ spaces the conditions for the ten operators $\subsub AH{}$, 
$\subsub AKi$, $\subsub APi$, and $\subsub AJi$ turn out to be 
\begin{eqnarray} 
&	\Lambda \biggl([H_0,\subsub AH{}]+H_I\subsub AH{}+H_I-\subsub
	AH{}H_I\subsub AH{}\biggr)\eta 
&=	0\label{bestH}\\
&	\Lambda \biggl([{K_0}_i,\subsub AKi]+{K_I}_i\subsub
	AKi+{K_I}_i-\subsub AKi{K_I}_i\subsub AKi\biggr)\eta
&=	0\label{bestK}\\
&	\Lambda[P_i,\subsub APi]\eta
&=	0\label{bestP}\\
&	\Lambda[J_i,\subsub AJi]\eta
&=	0\label{bestJ}
\end{eqnarray}
Here we made use of the assumption that ${\cal L}_I$ and hence $H_I$
and ${K_I}_i$ are linear in the meson operators such that $\eta
H_I\eta=0=\eta {K_I}_i\eta$. 
If one and the same $A$ can be found that fulfills the set of
conditions (\ref{bestH})-(\ref{bestJ}) the existence of ten effective
generators of the Poincar\'e group in the separate subspaces $\eta$
and $\Lambda$ is proven.
\section{Proof of the existence of $\fett A$}
The conditions (\ref{bestH})-(\ref{bestJ}) are nonlinear in the
$A$'s. One can linearize them by searching for $A$ in the form of a
Taylor expansion in the coupling constant
\begin{equation} 
A=\sum\limits_{\nu=1}^\infty \subsub A\nu{} g^\nu\label{reihe1}
\end{equation}
The term of order $g^0$ is absent since for a free theory
($g\rightarrow 0$) the generators are already blockdiagonal and ${\cal
U}=1$
is achieved with $A=0$. It is easy to rewrite the set under the
assumption (\ref{reihe1}) by equating equal powers of $g$. Keeping in
mind that we assumed $\eta H_I\eta=0=\eta{K_I}_i\eta$ we get the result:
\begin{eqnarray}
	{[}H_0,\subsub AH1]
&=&	-H_I\eta\label{bestH1}\\
	{[}H_0,\subsub AH2]
&=&	-\Lambda H_I\subsub AH1\label{bestH2}\\
	{[}H_0,\subsub AH{n+1}]
&=&	-\Lambda H_I\subsub AHn+\sum\limits_{\nu=1}^{n-1}
	\subsub AH\nu H_I\subsub AH{n\mbox -\nu}
	\hskip20pt n\ge 2\label{bestHn}\\
	{[}{K_0}_i,\subsub A{K_i}1]
&=&	-K_I\eta\label{bestK1}\\
	{[}{K_0}_i,\subsub A{K_i}2]
&=&	-\Lambda K_I\subsub A{K_i}1\label{bestK2}\\
	{[}{K_0}_i,\subsub A{K_i}{n+1}]
&=&	-\Lambda {K_I}_i\subsub A{K_i}n+\sum\limits_{\nu=1}^{n-1}
	\subsub A{K_i}\nu {K_I}_i\subsub A{K_i}{n\mbox -\nu}
	\hskip20pt n\ge 2\label{bestKn}\\
	{[}P_i,\subsub A{P_i}n]
&=&	0\hskip 20pt n\ge 1\label{bestPn}\\
	{[}J_i,\subsub A{J_i}n]
&=&	0\hskip 20pt n\ge 1\label{bestJn}
\end{eqnarray}
Let us introduce short hand notations. An arbitrary Fock state in the
$\Lambda$ space describing a certain number of noninteracting
particles with momenta $\fett p$ is simply denoted by
$|\Lambda\rangle$. Its energy, the eigenvalue to $H_0$, is denoted by
$E_\Lambda$. The projection operator into the $\Lambda$ space is
\begin{equation}
\Lambda\equiv\int\limits_\Lambda\dd p_\Lambda\
|\Lambda\rangle\langle\Lambda|\label{konv2}
\end{equation}
where $\ \dd p_\Lambda$ stands for all momentum integrations. Similarly
we denote an arbitrary state in the $\eta$ space by $|\eta\rangle$,
its energy by $E_\eta$ and the projection operator into the $\eta$ space
by
\begin{equation}
\eta\equiv\int\limits_\eta\dd p_\eta\
|\eta\rangle\langle\eta|\label{konv2a}
\end{equation}
Then it is very convenient to use the following notation
\begin{equation}
\int\limits_{\eta,\Lambda}\dd p_\Lambda\ \dd p_\eta{1\over E_\Lambda-E_\eta}
|\Lambda\rangle\langle\Lambda|B|\eta\rangle\langle\eta|
\equiv
{1\over E_\Lambda-E_\eta}\Lambda B\eta\label{konv3}
\end{equation}
where $B$ is an arbitrary operator. 

The set (\ref{bestH1})-(\ref{bestHn}) can now be solved
recursively. Note that according to Eq. (\ref{A}) $A$ connects the
$\Lambda$ and the $\eta$ spaces.
Using now the notation (\ref{konv3}) one finds the
following explicit expressions for $\subsub AHn$:
\begin{eqnarray}
\subsub AH1
&=&	-{1\over E_\Lambda-E_\eta}\Lambda H_I\eta\label{bestH1a}\\
\subsub AH2
&=&	-{1\over E_\Lambda-E_\eta}\Lambda H_I\subsub
	AH1\eta\label{bestH2a}\\
\subsub AH{n+1}
&=&	-{1\over E_\Lambda-E_\eta}\Lambda H_I\subsub AHn\eta
	+\sum\limits_{\nu=1}^{n-1}{1\over E_\Lambda-E_\eta}\Lambda 
	\subsub AH\nu H_I\subsub AH{n-\nu}\eta\hskip20pt n\ge2
	\label{bestHna}
\end{eqnarray}
With the help of the Lie algebra (\ref{lie1})-(\ref{lie9}) we will now
show by induction that $\subsub AHn$ given by Eqs.
(\ref{bestH1a})-(\ref{bestHna}) satisfies also Eqs.
(\ref{bestK1})-(\ref{bestJn}) and therefore one and the same $A$
blockdiagonalizes all ten generators.

First we look at Eq. (\ref{bestPn}). Because of 
\begin{equation}
[P_i,\Lambda]=[P_i,\eta]=0 \label{PLambda}
\end{equation}
one has the following identity for any operator $B(H)$
\begin{equation}
[P_i,{1\over E_\Lambda-E_\eta}\Lambda B(H)\eta]={1\over
E_\Lambda-E_\eta}\Lambda[P_i,B(H)]\eta\label{PmitE}
\end{equation}
Further, since $P_i$ commutes with $H_0$ and $H$, it also commutes
with $H_I\equiv H-H_0$. Consequently we get
\begin{eqnarray}
[P_i,\subsub AH1]&=&
	-[P_i,{1\over E_\Lambda-E_\eta}\Lambda H_I\eta]\nonumber\\
&=&	-{1\over E_\Lambda-E_\eta}\Lambda[P_i,H_I]\eta=0\label{bewP1}
\end{eqnarray}
By induction this carries over to $\subsub AH2$ and $\subsub AHn$ with
$n\ge 3$ and we can write
\begin{equation}
\subsub APi=\subsub AH{}\label{apgleichah}
\end{equation}
The proof of (\ref{bestJn}) using $\subsub AH{}$ is very
similar. Eqs. (\ref{PLambda})-(\ref{bewP1}) remain valid replacing
$P_i$ by $J_i$ and $J_i$ commutes with $H$ and $H_0$. Consequently we
can also put
\begin{equation}
\subsub AJi=\subsub AH{}\label{ajgleichah}
\end{equation}
The proof that $\subsub AHn$ solves the set
(\ref{bestK1})-(\ref{bestKn}) is the difficult one. We need the
following relations. From
\begin{equation}
{[}H,K_i]=-\ii P_i\label{hmitk}
\end{equation}
and the linear dependence of $H$ and $K_i$ on $g$ (see Eqs.
(\ref{HH}), (\ref{KK}), and (\ref{HHa})) we find easily by equating operators
related to equal powers in $g$
\begin{eqnarray} 
{[}H_0,{K_0}_i]&=&-\ii P_i\label{H0mitK0}\\
{[}H_0,{K_I}_i]&=&{[}{K_0}_i,H_I]\label{HImitK0}\\
{[}H_I,{K_I}_i]&=&0\label{HImitKI}
\end{eqnarray}
Further one has
\begin{equation}
{[}{K_0}_i,C(H_0)]=\ii P_i{\partial\over\partial H_0}C(H_0)\label{KmitH0}
\end{equation}
where $C$ depends on $H_0$ only. Because of the free state kinematics
it is also easily seen that
\begin{equation}
{[}{K_0}_i,\Lambda]={[}{K_0}_i,\eta]=0
\end{equation}
Using all that one verifies that
\begin{eqnarray}
\lefteqn{{[}{K_0}_i,{1\over E_\Lambda-E_\eta}\Lambda B(H)\eta]}\nonumber\\
&=&{[}{K_0}_i,{1\over H_0-E_\eta}\Lambda B(H)\eta]\nonumber\\
&=&{[}{K_0}_i,{1\over H_0-E_\eta}]\Lambda B\eta+
{1\over E_\Lambda-E_\eta}\Lambda{[}{K_0}_i,B]\eta+
\Lambda B{[}{K_0}_i,{1\over E_\Lambda-H_0}]\eta\\
&=&{1\over E_\Lambda-E_\eta}\Lambda{[}{K_0}_i,B]\eta\label{KmitE}
\end{eqnarray}
This enables us now to show that $\subsub AH1$ solves Eq. (\ref{bestK1}):
\begin{eqnarray}
\lefteqn{{[}{K_0}_i,\subsub AH1]}\nonumber\\
&=&	{[}{K_0}_i,-{1\over E_\Lambda-E_\eta}\Lambda
	H_I\eta]\nonumber\\
&=&	-{1\over E_\Lambda-E_\eta}\Lambda{[}{K_0}_i,H_I]\eta\nonumber\\
&=&	-{1\over
	E_\Lambda-E_\eta}\Lambda{[}H_0,{K_I}_i]\eta\nonumber\\
&=&	-{1\over
	E_\Lambda-E_\eta}\Lambda\Bigl(E_\Lambda-E_\eta\Bigr){K_I}_i\eta
=	-\Lambda {K_I}_i\eta\label{indK1}
\end{eqnarray}
Next let us verify that $\subsub AH2$ from Eq. (\ref{bestH2a})
solves Eq. (\ref{bestK2}):
\begin{eqnarray}
\lefteqn{{[}{K_0}_i,\subsub AH2]=
	-{[}{K_0}_i,{1\over E_\Lambda-E_\eta}\Lambda H_I\subsub
	AH1\eta]}
	\nonumber\\
&=&	-{1\over E_\Lambda-E_\eta}\Lambda{[}{K_0}_i,H_I\subsub
	AH1]\eta\nonumber\\
&=&	-{1\over E_\Lambda-E_\eta}\Lambda{[}H_0,{K_I}_i]\subsub
	AH1\eta\nonumber\\
&&	-{1\over E_\Lambda-E_\eta}\Lambda H_I{[}{K_0}_i,\subsub
	AH1]\eta\label{indK2a}
\end{eqnarray}
Using Eq. (\ref{indK1}) and applying $H_0$ this can be rewritten as
\begin{eqnarray}
{[}{K_0}_i,\subsub AH2]&=&-{1\over E_\Lambda-E_\eta}\Lambda{K_I}_i
	(E_\Lambda-H_0)\subsub AH1\eta\nonumber\\
&&	+{1\over E_\Lambda-E_\eta}\Lambda H_I{K_I}_i\eta
\label{indK2b}
\end{eqnarray}
Further using Eq. (\ref{HImitKI}) we get
\begin{eqnarray}
\lefteqn{{[}{K_0}_i,\subsub AH2]=-{1\over
	E_\Lambda-E_\eta}\Lambda{K_I}_i\Bigl((E_\Lambda-H_0)\subsub
	AH1-H_I\Bigr)\eta}\nonumber\\
&=&	-{1\over
	E_\Lambda-E_\eta}\Lambda{K_I}_i\Bigl(E_\Lambda\subsub AH1
	-{[}H_0,\subsub AH1]-\subsub AH1H_0-H_I\Bigr)\eta\nonumber\\
&=&	-\Lambda{K_I}_i\subsub AH1-{1\over
	E_\Lambda-E_\eta}\Lambda{K_I}_i\Bigl(
	-{[}H_0,\subsub AH1]-H_I\Bigr)\eta\nonumber\\
&=&	-\Lambda{K_I}_i\subsub AH1\label{indK2}
\end{eqnarray}
The last step follows from Eq. (\ref{bestH1}) and proves $\subsub
AH2=\subsub AK2$.

Due to the structure of the set (\ref{bestK1})-(\ref{bestKn}) it turns out
that the proof for $\subsub AH3$ and $\subsub AH4$ should also be
treated separately. Since the algebra is rather lengthy and the steps
used for the general case $\subsub AHn$, $n\ge5$ include all the ones
for the simpler cases $n=3$ and $n=4$ we leave the verification for
those simpler cases to the reader.

We embark now in the proof that $\subsub AH{n+1}$, $n\ge4$ is a
solution of Eq. (\ref{bestKn}) provided every
$\subsub AH\nu$, $\nu\le n$ is a solution of the set
(\ref{bestK1})-(\ref{bestKn}). From (\ref{bestHna}) we get
\begin{equation} 
{[}{K_0}_i\,,\subsub AH{n+1}]
={1\over E_\Lambda-E_\eta}\Lambda\Biggl[{K_0}_i\,,\biggl(-H_I\subsub AHn
+\sum\limits_{\nu=1}^{n-1}\subsub AH\nu H_I\subsub
AH{n-\nu}\biggr)\Biggr]\eta\label{schritt1}
\end{equation}
Using Eq. (\ref{HImitK0}) this can be rewritten as
\begin{eqnarray}
\lefteqn{{[}{K_0}_i\,,\subsub AH{n+1}]}\nonumber\\
&=&	{1\over E_\Lambda-E_\eta}\Lambda
\Biggl(
	-\left[H_0\,,{{ K}_I}_i\right]\subsub AHn
	-H_I\left[{{ K}_0}_i\,,\subsub AHn\right]\nonumber\\
&&	+\sum\limits_{\nu =1}^{n-1}
	\left[{{ K}_0}_i\,,\subsub AH\nu\right]
	H_I\subsub AH{n\mbox -\nu}\nonumber\\
&&	+\sum\limits_{\nu =1}^{n-1}
	\subsub AH\nu\left[H_0\,,{{ K}_I}_i\right]
	\subsub AH{n\mbox -\nu}\nonumber\\
&&	+\sum\limits_{\nu =1}^{n-1}
	\subsub AH\nu H_I
	\left[{{ K}_0}_i\,,\subsub AH{n\mbox -\nu}\right]\Biggr)\eta
\label{schritt2}
\end{eqnarray}
The first term in Eq. (\ref{schritt2}) is changed as
\begin{equation}
-{1\over E_{\Lambda}-E_{\eta}}\Lambda
	\left[H_0\,,{{  K}_I}_i\right]\subsub AHn\eta
=	-\Lambda{{  K}_I}_i\subsub AHn\eta
	-{1\over E_{\Lambda}-E_{\eta}}\Lambda
	{{  K}_I}_i\left(E_{\eta}-H_0\right)\subsub AHn\eta
\label{schritt3}
\end{equation}
Inserting this into Eq. (\ref{schritt2}) and using the assumption that
$\subsub AH\nu$, $\nu\le n$ solves the set
(\ref{bestK1})-(\ref{bestKn}) we get
\begin{eqnarray}
{[}{{  K}_0}_i\,,\subsub AH{n+1}]
&=&	 -\Lambda{{  K}_I}_i\subsub AHn\nonumber\\
&&	+{1\over E_{\Lambda}-E_{\eta}}\Lambda
	\Biggl(
	-{{  K}_I}_i\left(E_\eta-H_0\right)\subsub AHn\nonumber\\
&&	+\biggl[
	 H_I\Lambda'{{  K}_I}_i\subsub AH{n\mbox -1}
	 -H_I\sum\limits_{\nu=1}^{n-2}\subsub AH\nu{{  K}_I}_i
	 \subsub AH{n\mbox -1\mbox -\nu}
	  \biggr]\nonumber\\
&&	+\biggl[
	 -{{  K}_I}_i\eta' H_I\subsub AH{n\mbox -1}
	 -{{  K}_I}_i\subsub AH1 H_I\subsub AH{n\mbox -2} \nonumber\\
&&	+\sum\limits_{\nu=3}^{n-1}\biggl(
	-{{  K}_I}_i\subsub AH{\nu\mbox -1}
	 +\sum\limits_{\nu'=1}^{\nu-2}\subsub AH{\nu'}
	{{  K}_I}_i\subsub AH{\nu\mbox -1\mbox -\nu'}
	 \biggr)H_I\subsub AH{n\mbox -\nu}
	 \biggr]\nonumber\\
&&	+\sum\limits_{\nu =1}^{n-1}
	\subsub AH\nu\left[H_0\,,\,{{  K}_I}_i\right]
	\subsub AH{n\mbox -\nu}\nonumber\\
&&	+\biggl[
	 \sum\limits_{\nu =1}^{n-3}
	 \subsub AH\nu H_I\biggl(
	 -{{  K}_I}_i\subsub AH{n\mbox -\nu\mbox -1}
	 +\sum\limits_{\nu' =1}^{n-\nu-2}\subsub AH{\nu'} {{  K}_I}_i
	 \subsub AH{n\mbox -\nu\mbox -1\mbox -\nu'}
	 \biggr)\nonumber\\
&&	-\subsub AH{n\mbox -2} H_I{{  K}_I}_i\subsub AH1
	 -\subsub AH{n\mbox -1} H_I{{  K}_I}_i
	 \biggr]\ \Biggr)\eta\nonumber\\
&&	+{1\over E_{\Lambda}-E_{\eta}}\Lambda\Bigl(
	{{  K}_I}_i\Lambda' H_I\subsub AH{n\mbox -1}
	 -{{  K}_I}_i\Lambda' H_I\subsub AH{n\mbox -1}
	 \Bigr)\eta\label{schritt4}
\end{eqnarray}
The square brackets are inserted to group the terms together resulting
from the commutators with ${K_0}_i$. Moreover we added a zero at the
end and used the identity
\begin{equation}
\subsub AHnH_I\Lambda{K_I}_i=\subsub AHnH_I{K_I}_i\label{schritt4a}
\end{equation}
which is valid because $\subsub AH{}$ has an $\eta$-projector on the
right. Eq. (\ref{schritt4}) can be simplified by means of the two
identities
\begin{equation}
  -{{ K}_I}_i\Lambda H_I
  +H_I\Lambda{{ K}_I}_i
  -{{ K}_I}_i\eta H_I
=-{{ K}_I}_iH_I
  +H_I\Lambda{{ K}_I}_i
=-H_I\eta{{ K}_I}_i\label{schritt5}
\end{equation}
and
\begin{eqnarray}
{1\over E_{\Lambda}-E_{\eta}}\Lambda
	\Bigl(
	 {{ K}_I}_i(H_0-E_{\eta})\subsub AHn
	 +{{ K}_I}_i\Lambda' H_I\subsub AH{n\mbox -1}
	 -{{ K}_I}_i\sum\limits_{\nu=3}^{n-1}
	 \subsub AH{\nu\mbox -1} H_I
	 \subsub AH{n\mbox -\nu}\nonumber\\
	-{{ K}_I}_i\subsub AH1 H_I\subsub AH{n\mbox -2}
	\Bigr)\eta=0\label{schritt6}
\end{eqnarray}
The second one is just a consequence of Eq. (\ref{bestHn}). Using
Eqs. (\ref{schritt5}) and (\ref{schritt6}) we find
\begin{eqnarray}
{[}{{ K}_0}_i\,,\subsub AH{n+1}]
&=&	-\Lambda{{ K}_I}_i\subsub AHn\nonumber\\
&&	+{1\over E_{\Lambda}-E_{\eta}}\Lambda
	\Biggl(
	 -H_I\eta'{{ K}_I}_i\subsub AH{n-1}-H_I
	 \sum\limits_{\nu=1}^{n-2}\subsub AH\nu{{ K}_I}_i
	 \subsub AH{n\mbox -1\mbox -\nu}\nonumber\\
&&	+\sum\limits_{\nu=3}^{n-1}
	\sum\limits_{\nu'=1}^{\nu-2}
	 \subsub AH{\nu'}{{ K}_I}_i\subsub AH{\nu\mbox -1\mbox -\nu'}
	H_I\subsub AH{n\mbox -\nu}\nonumber\\
&&	+\sum\limits_{\nu=1}^{n-1}
	 \subsub AH{\nu}{[}H_0\,,{{ K}_I}_i]\subsub AH{n\mbox -\nu}\nonumber\\
&&	+\sum\limits_{\nu=1}^{n-3}\subsub AH{\nu} H_I
	\Bigl(-{{ K}_I}_i\subsub AH{n\mbox -\nu\mbox -1}
	+\sum\limits_{\nu'=1}^{n-\nu-2}
	 \subsub AH{\nu'}{{ K}_I}_i
	\subsub AH{n\mbox -\nu\mbox -1\mbox -\nu'}\Bigr)\nonumber\\
&&	-\subsub AH{n\mbox -2} H_I{{ K}_I}_i\subsub AH1
	-\subsub AH{n\mbox -1} H_I{{ K}_I}_i
	\Biggr)\eta\label{schritt7}
\end{eqnarray}
Next we exchange the orders of summation
\begin{eqnarray}
\lefteqn{\sum\limits_{\nu=3}^{n-1}
	 \sum\limits_{\nu'=1}^{\nu-2}
	   \subsub AH{\nu'}{{K}_I}_i\subsub AH{\nu\mbox -1\mbox -\nu'}
	 H_I\subsub AH{n\mbox -\nu}}\nonumber\\
&=&	\sum\limits_{\nu'=1}^{n-3}
	\sum\limits_{\nu=1}^{n-\nu'-2}
	\subsub AH{\nu'}{{K}_I}_i\subsub AH\nu
	H_I\subsub AH{n\mbox -\nu'\mbox -1\mbox -\nu}\label{stop5}
\end{eqnarray}
and
\begin{eqnarray}
\lefteqn{\sum\limits_{\nu=1}^{n-3}
	 \subsub AH\nu H_I
	 \sum\limits_{\nu'=1}^{n-\nu-2}
	  \subsub AH{\nu'}{{K}_I}_i
	 \subsub AH{n\mbox -\nu\mbox -1\mbox -\nu'}}\nonumber\\
&=&	\sum\limits_{\nu=1}^{n-3}
	\subsub AH\nu H_I
	 \sum\limits_{\nu'=1+\nu}^{n-2}
	\subsub AH{\nu'\mbox -\nu}{K_I}_i
	\subsub AH{n\mbox -1\mbox -\nu'}\nonumber\\
&=&	\sum\limits_{\nu'=2}^{n-2}
	\sum\limits_{\nu=1}^{\nu'-1}
	\subsub AH\nu H_I
	\subsub AH{\nu'\mbox -\nu}{{K}_I}_i
	\subsub AH{n\mbox -1\mbox -\nu'}\label{stop6}
\end{eqnarray}
Using Eq. (\ref{stop5}) we can group together some terms from
Eq. (\ref{schritt7}) taking Eq. (\ref{HImitKI}) and the set
(\ref{bestH1})-(\ref{bestHn}) into account:
\begin{eqnarray}
\lefteqn{\sum\limits_{\nu=3}^{n-1}
	\sum\limits_{\nu'=1}^{\nu-2}
	 \subsub AH{\nu'}{{ K}_I}_i\subsub AH{\nu\mbox -1\mbox -\nu'}
	H_I\subsub AH{n\mbox -\nu}
	  -\sum\limits_{\nu'=1}^{n-3}
	 \subsub AH{\nu'}{{K}_I}_i H_I 
	 \subsub AH{n\mbox -\nu'\mbox -1}}\nonumber\\
&&	{-\subsub AH{n\mbox -2}{{K}_I}_i H_I\subsub AH1
	 -\subsub AH{n\mbox -1}{{K}_I}_i H_I}\nonumber\\
&=&	\sum\limits_{\nu'=1}^{n-3}
	 \sum\limits_{\nu=1}^{n-\nu'-2}
	  \subsub AH{\nu'}{{K}_I}_i\subsub AH\nu
	 H_I\subsub AH{n\mbox -\nu'\mbox -1\mbox -\nu}
	  -\sum\limits_{\nu'=1}^{n-3}
	 \subsub AH{\nu'}{{K}_I}_i H_I 
	 \subsub AH{n\mbox -\nu'\mbox -1}\nonumber\\
&&	 {-\subsub AH{n\mbox -2}{{K}_I}_i H_I\subsub AH1
	 -\subsub AH{n\mbox -1}{{K}_I}_i H_I}\nonumber\\
&=&	 +\sum\limits_{\nu'=1}^{n-3}
	\subsub AH{\nu'}{{K}_I}_i{[}H_0,\subsub AH{n\mbox -\nu'}]
	\nonumber\\
&&	+\subsub AH{n\mbox -2}{{K}_I}_i{[}H_0,\subsub AH2]
	+\subsub AH{n\mbox -1}{{K}_I}_i{[}H_0,\subsub AH1]\nonumber\\
&=&	\sum\limits_{\nu'=1}^{n-1}
	\subsub AH{\nu'}{{K}_I}_i{[}H_0,\subsub AH{n\mbox -\nu'}]
   \label{schritt8}
\end{eqnarray}
Similarly the expression (\ref{stop6}) can
be grouped together with two more terms from Eq. (\ref{schritt7})
\begin{eqnarray}
\lefteqn{\sum\limits_{\nu=1}^{n-3}
	 \subsub AH\nu H_I
	 \sum\limits_{\nu'=1}^{n-\nu-2}
	  \subsub AH{\nu'}{{K}_I}_i
	 \subsub AH{n\mbox -\nu\mbox -1\mbox -\nu'}
	 -\Lambda H_I\eta'{{K}_I}_i\subsub AH{n\mbox -1}
	 -\Lambda H_I
	 \sum\limits_{\nu'=1}^{n-2}
	\subsub AH{\nu'}{{K}_I}_i\subsub AH{n\mbox -1\mbox -\nu'}}
	\nonumber\\
&=&\!\!\!\!
	 \sum\limits_{\nu'=2}^{n-2}
	 \Bigl(-\Lambda H_I\subsub AH{\nu'}\!
	 +\!\sum\limits_{\nu=1}^{\nu'-1}\!
	 \subsub AH\nu H_I
	 \subsub AH{\nu'\mbox -\nu}\Bigr)
	{{K}_I}_i\subsub AH{n\mbox -1\mbox -\nu'}
	 -\Lambda H_I\eta'{{K}_I}_i\Lambda\subsub AH{n\mbox -1}\nonumber\\
&&	\hskip250pt   
	 -\Lambda H_I\subsub AH1{{K}_I}_i\subsub AH{n\mbox -2}
	 \nonumber\\
&=&\!\!\!\!
	\sum\limits_{\nu'=2}^{n-2}
	 {[}H_0,\subsub AH{\nu'\mbox {\tiny +}1}]
	 {{K}_I}_i\subsub AH{n\mbox -1\mbox -\nu'}
	 \!+\!{[}H_0,\subsub AH1]
	 {{K}_I}_i\subsub AH{n\mbox -1}
	  \!\!+\!{[}H_0,\subsub AH2]{{K}_I}_i\subsub AH{n\mbox -2}
	  \nonumber\\
&=&	\sum\limits_{\nu'=3}^{n-1}
	 {[}H_0,\subsub AH{\nu'}]
	 {{K}_I}_i\subsub AH{n\mbox -\nu'}
	 +{[}H_0,\subsub AH1]
	 {{K}_I}_i\subsub AH{n\mbox -1}
	 +{[}H_0,\subsub AH2]{{K}_I}_i\subsub AH{n\mbox -2}
	 \nonumber\\
&=&	\sum\limits_{\nu'=1}^{n-1}
	{[}H_0,\subsub AH{\nu'}]
	 {{K}_I}_i\subsub AH{n\mbox -\nu'}\label{schritt9}
\end{eqnarray}
Again we used the set (\ref{bestH1})-(\ref{bestHn}) several times. Inserting
all that into Eq.
(\ref{schritt7}) that expression simplifies greatly and leads to the
desired result:
\begin{eqnarray}
{[}{{ K}_0}_i\,,\subsub AH{n+1}]
&=&
	 -\Lambda{{ K}_I}_i\subsub AHn\nonumber\\
&&	+{1\over E_{\Lambda}-E_{\eta}}\Lambda
	\sum\limits_{\nu=1}^{n-1}
	\Biggl[
	\subsub AH\nu{{ K}_I}_i{[}H_0\,,\subsub AH{n\mbox -\nu}]
	+\subsub AH\nu{[}H_0\,,{{ K}_I}_i]\subsub AH{n\mbox -\nu}
	\nonumber\\
&&	\hskip200pt
	+{[}H_0\,,\subsub AH\nu] {{ K}_I}_i\subsub AH{n\mbox -\nu}
	\Biggr]
	\eta\nonumber\\
&=&
	 -\Lambda{{ K}_I}_i\subsub AHn\nonumber\\
&&	+{1\over E_{\Lambda}-E_{\eta}}\Lambda
   \sum\limits_{\nu=1}^{n-1}
   \subsub AH\nu\eta'{{ K}_I}_i\Lambda'\subsub AH{n\mbox -\nu}
   (E_{\Lambda'}-E_{\eta}+E_{\eta'}-E_{\Lambda'}+E_{\Lambda}-E_{\eta'})
   \eta\nonumber\\
&=&-\Lambda{{ K}_I}_i\subsub AHn
   +\sum\limits_{\nu=1}^{n-1}
   \subsub AH\nu{{ K}_I}_i\subsub AH{n\mbox -\nu}
   \label{schritt10}
\end{eqnarray}
This is our final result stating that $\subsub AH{n+1}$ solves Eq.
(\ref{bestKn}) for $n\ge4$ provided that $\subsub AH\nu$, $\nu=
1,2,3,4$ is a solution of Eqs. (\ref{bestK1})-(\ref{bestKn}). This however
is the induction assumption and has been shown before.

We conclude that
\begin{equation}
\subsub A{K_i}n=\subsub AHn\hskip20pt n\ge1\label{erg1}
\end{equation}
and hence
\begin{equation}
\subsub A{K_i}{}=\subsub AH{}
\label{erg2}
\end{equation}
Together with Eqs. (\ref{apgleichah}) and (\ref{ajgleichah}) we arrive at
the important result that $\subsub AH{}$ solves the set of all
Eqs. (\ref{bestH})-(\ref{bestJ}) and all indices on the $A$'s can
be omitted. In practice one will use the recursion relations
(\ref{bestH1a})-(\ref{bestHna}) for calculating $A$ since they are
easier to solve than the ones resulting from Eqs.
(\ref{bestK1})-(\ref{bestKn}) and further the conditions (\ref{bestPn})
and (\ref{bestJn}) are not specific enough.

\section{Summary}
The ten generators of the Poincar\'e group acting in the full Fock
space of nucleons, antinucleons, and mesons and having the form
(\ref{HH})-(\ref{JJ}) according to often used Lagrange densities can
be blockdiagonalized at the same time by a single unitary
transformation. Thereby the blocks are defined by two projection
operators which span the Fock space one refering to a fixed
number of nucleons and the other to the rest of the space. As a
consequence the resulting unitarily transformed generators act in these
two spaces separately and specifically we gained effective generators
in the space of $N$ nucleons which are true representations of the
Poincar\'e algebra. This result has been proven using a (formal) power
series expansion in the coupling constant. We see the importance of
that result in the existence proof. Clearly in practice this series
has to be truncated as is usually done in evaluating $NN$ forces in
low orders of meson exchanges.

Results of numerical studies in
\cite{muller2} - \cite{hagl} are promising. They show for
instance that the relativistic energy-momentum relation of a two body
state is rather well fulfilled if one solves the Schr\"odinger
equation using the effective Hamiltonian in frames where the total
momentum of the two-body system is different from zero. They also show
that contributions to the relativistic Hamiltonian which remain
undetermined in the scheme of an $1\over{\rm c}^2$ expansion of the
Poincar\'e generators \cite{folkra} can now be determined for any given
field theory and are different from zero. In fact in the numerical
examples studied \cite{muller2} they are as important as those
enforced by the Poincar\'e algebra. Thus that scheme discussed in this
article provides interesting structural insight.

In addition one can pose now various questions.
One of the effective generators,
the Hamiltonian in the space of $N$ nucleons, will contain $NN$ but
also many body forces. Will they fulfill the cluster separability?
Since the effective generators are constructed in a power series
expansion in the coupling constant one encounters in all orders $g^n$
with $n\ge4$ meson exchange diagrams together with vertex corrections
for instance. The question then arises whether the Poincar\'e algebra
for the ten effective generators, which is fulfilled in each order in
$g$, requires all the terms of a certain order in $g$ or whether
subgroups fulfill the algebra separately. In the first case the
Poincar\'e algebra would impose conditions on the acceptable vertex
corrections which would play the role of strong formfactors. Further
investigations of that type are planned.

\end{document}